\documentclass[crop]{CSL}

\jname{International Journal of Astrobiology}%

\usepackage{graphicx}	
\usepackage{amsmath}	
\usepackage{natbib}
\newcommand{\beq}{\begin{equation}}
\newcommand{\eeq}{\end{equation}}
\newcommand{\bea}{\begin{eqnarray}}
\newcommand{\eea}{\end{eqnarray}}

\newcommand{\FP}{\text{FP}}
\newcommand{\FN}{\text{FN}}

\newcommand{\ntot}{N_\text{tot}}
\newcommand{\mtot}{M_\text{tot}}
\newcommand{\ndet}{N_\text{det}}
\newcommand{\Mdet}{M_\text{det}}

\newcommand{\tu}{\textnormal}
\newcommand{\ctot}{c_\text{tot}}
\newcommand{\qtot}{q_\text{tot}}
\newcommand{\tdev}{t_\text{dev}}

\newcommand{\fbio}{f_\text{bio}}

\begin{document}

\supertitle{Research Paper}

\title[Information Gain for Biosignature Missions]{Information Gain as a Tool for Assessing Biosignature Missions}

\author[Fields, Gupta and Sandora]{Benjamin Fields$^{1,2\dagger}$, 
        Sohom Gupta$^{1,3\dagger}$,
        McCullen Sandora$^{1*}$}

\address{\add{}{${}^\dagger$Equal contribution} \add{1}{Blue Marble Space Institute of Science} \add{2}{Wheaton College} and \add{3}{Indian Institute of Science Education and Research Kolkata}}

\corres{\name{McCullen Sandora} \email{mccullen@bmsis.org}}

\begin{abstract}
We propose the mathematical notion of \emph{information gain} as a way of quantitatively assessing the value of biosignature missions.  This makes it simple to determine how mission value depends on design parameters, prior knowledge, and input assumptions.  We demonstrate the utility of this framework by applying it to a plethora of case examples: the minimal number of samples needed to determine a trend in the occurrence rate of a signal as a function of an environmental variable, and how much cost should be allocated to each class of object; the relative impact of false positives and false negatives, with applications to Enceladus data and how best to combine two signals; the optimum tradeoff between resolution and coverage in the search for lurkers or other spatially restricted signals, with application to our current state of knowledge for solar system bodies; the best way to deduce a habitability boundary; the optimal amount of money to spend on different mission aspects; when to include an additional instrument on a mission; the optimal mission lifetime; and when to follow/challenge the predictions of a habitability model.  In each case, we generate concrete, quantitative recommendations for optimising mission design, mission selection, and/or target selection.
\end{abstract}

\keywords{biosignatures, statistics, astrobiology}

\selfcitation{}

\received{xx xxxx xxxx}

\revised{xx xxxx xxxx}

\accepted{xx xxxx xxxx}

\maketitle

\Fpagebreak

\section{Introduction}

The overarching goal of the field of astrobiology is to search for signs of life elsewhere in the universe.  As it is a highly interdisciplinary endeavour, the strategies proposed to achieve this are multifaceted, spanning a range of solar system and exoplanet targets and detection techniques.  As such, it is sometimes challenging even for seasoned practitioners to assess the value of upcoming missions.  Recently, there have been efforts to address this.  The Ladder of Life Detection \citep{neveu2018ladder}, Confidence of Life Detection score \citep{green2021call}, and Standards of Evidence framework \citep{meadows2022community} aim to provide a language for communicating the evidential strength of any observed biosignature, both to the public and amongst scientists.  The nine axes of merit has been proposed as a framework for describing the utility of any technosignature detection mission in \citep{sheikh2020nine}.  \citep{lorenz2019calculating} provides a risk versus payoff framework for life detection missions, which even factors in the mission success probability.

While many of these frameworks are fantastic at isolating the relevant mission aspects necessary for judging the merits of a mission, they are fundamentally descriptive processes, and leave up to the judgement of the practitioner how to prioritise each mission dimension.  As such, they are not suited for any final decision making process, whether it be selecting between missions or optimising mission design.

In this note we propose a simple method of assessing mission value in terms of \emph{information gain}, and demonstrate the utility of this formalism through considering several idealised mission designs.  This framework allows to distil the specifics of a mission down to a single number, which is useful for many purposes.  At the outset, we should acknowledge the shortcomings of this procedure: clearly, there is no unique way of ranking missions; our approach is capable of providing one such quantification, but other factors are surely important.  While our approach is encompasses many mission aspects, it specifically disregards, for instance, any ancillary benefits a mission would provide apart from biosignature detection.  Secondly, we wholeheartedly disavow mistaking performance metrics for the indefinite concept of mission value.  Any practitioner interested in using a method like this should be well aware of the dangers inherent in using performance metrics for decision making- this is exemplified in Goodhart's law, whereby when a quantitative proxy is used to drive decision making, there is a tendency for people to optimise for the proxy itself, sometimes to the detriment of the original intended goal \citep{chrystal2003goodhart}.  As such, we do not endorse a plan to `maximise information gain', nor do we believe it would be particularly useful to very precisely calculate the information of any particular mission.  The intent of our framework is to provide a rough heuristic that is easily calculable in a manner that makes the dependence on mission parameters highly apparent, so that these can be quickly and straightforwardly varied to determine their effect on the overall mission value.

In the remainder of this section we go into more detail on the process of calculating information gain for a given mission setup.  In the following sections, we demonstrate the utility and flexibility of this approach by applying it to several disparate scenarios, and deriving concrete, quantitative recommendations for optimising mission design.  In section \ref{section_ratio} we apply this formalism to the challenge of determining whether the occurrence rate of a biosignature depends on some system variable.  In section \ref{section_tradeoff} we consider the effects false positives and false negatives have on information.  In section \ref{section_lurkers} we consider a tradeoff between resolution and coverage in searching for a biosignature.  Section \ref{section_further} contains more applications, including: how to best determine the limiting value of some system parameter for life, how to apportion budget to various aspects of a mission, when to include an instrument on a mission, how to determine the optimal mission lifetime, and when to follow a habitability model, and when to challenge it.

\subsection{Bayes Estimation and Information gain}

In this work we propose to operationalize the dictum of many proposed missions, to ``maximise science return'', in the following way- by equating science return with \emph{information gain}, in the information theoretic sense, we may write the science return as
\beq
S=\int df\, p(f) \,\log p(f)\label{S}
\eeq
This explicitly incorporates a value for how to reduce raw mission data, in the form of a bit stream, down to the distilled representation relevant for our immediate purposes.  As such, this framework is more general, and can suitably be applied to a variety of scientific goals across disciplines.

As a simple application of the above framework, let us imagine the simplest setup, with a single biosignature, in the absence of false positives and negatives, and drawing from a uniform population.  We wish to estimate the fraction of locations possessing this biosignature $f$, given a total number of systems surveyed by a mission $\ntot$, which returns $\ndet$ detections.  If the occurrence rate were known, the number of detections would follow a Bernoulli distribution, and so when the number of detections is known instead, the signal occurrence rate $f$ follows the conjugate distribution $f\sim \beta(\ntot, \ndet,f)$, where $\beta$ refers to the Beta distribution\footnote{This is dependent on the choice of the prior distribution for $f$, which we take here to be uniform. However, using other priors does not change the behaviour of the population mean and variance by much\citep{Sandora_2020}.} \citep{Sandora_2020}. 
\bea
\beta(\ntot, \ndet,f)&=&B(\ntot,\ndet)\,f^{\ndet}(1-f)^{\ntot-\ndet}\nonumber\\
B(\ntot,\ndet)&=&\frac{(\ntot+1)!}{\ndet!(\ntot-\ndet)!}
\eea
This distribution has the following mean and variance:
\beq
\mu = \frac{\ndet+1}{\ntot+2},\hspace{0.1 in} \sigma^2 = \frac{(\ndet+1)(\ntot-\ndet+1)}{(\ntot+2)^2(\ntot+3)}
\eeq

If the mean is well separated from 0, this has $\mu\rightarrow f_o=\ndet/\ntot$ and $\sigma^2\rightarrow f_o(1-f_o)/\ntot$.  If the mean is consistent with 0 to observed uncertainty, $\mu\rightarrow1/\ntot$ and $\sigma^2\rightarrow1/\ntot^2$.  While simple, already this setup indicates how precision depends on survey size, for multiple different regimes.  This can be related to science yield (information gain) through eqn. (\ref{S}).

For a broad class of distributions, including the setup above, information reduces to $S\approx \log\sigma$, so maximising information gain becomes equivalent to minimising variance, which sets the measurement uncertainty.  In this limit, 1 bit of information is gained if the error bars for the measured quantity are decreased by a factor of 2.

The above expression assumes initial ignorance.  For the case where we are updating knowledge obtained from a previous mission, we can instead use the Kullback-Leibler (KL) divergence as the generalisation: 
\beq
\Delta S = KL(p||q) = \int df\,p(f)\,\log\left(\frac{p(f)}{q(f)}\right)
\eeq

This is a measure of information gain as one moves from an initial distribution $p$ to a final one $q$. This has the property that $KL(p||q) \geq 0$, with equality only if $p=q$, guaranteeing that information is always gained with new missions.  This reduces to the prior case if we take the initial distribution $q$ to be uniform.

In the case where we are updating our knowledge of biosignature occurrence rate from a previous mission, both subject to the conditions outlined above, we have $p\sim \beta(\ntot,\ndet,f)$ and $q\sim\beta(\ntot^{(0)},\ndet^{(0)},f)$ then we can find the exact formula for information gain from \citep{RAUBER2008637}:
\bea
\Delta S&=&\log\frac{B\left(\ntot,\ndet\right)}{B\left(\ntot^{(0)},\ndet^{(0)}\right)}\nonumber\\
&&+\Delta (\ntot-\ndet)\psi(\ntot-\ndet+1)\nonumber\\
&&+\Delta \ndet\psi(\ndet+1)-\Delta\ntot\psi(\ntot+2)
\eea

where $\psi(x)=\frac{d\Gamma(x)}{dx}/\Gamma(x)$ is the digamma function. Using the approximation
$\psi(x) \sim \log x - \frac{1}{2x}$, we can see that this is a generalisation of the formula $\Delta S\approx \log\sigma$ to situations where we start with some information.  Additionally, in the limit where the sample size is incremented by 1 in an updated experiment, we find $\Delta S\approx1/\ntot$.

This framework can be compared to other, more rigorous statistical methods of estimating survey efficacy.  If one had data from a set of observations, the usual procedure would be to perform a battery of statistical tests (eg, Neyman Pearson, binomial, kernel tests) aimed at determining the precise degree to which the tested hypothesis is (dis)favoured compared to some null hypothesis.  While rejecting the null hypotheses to some specified confidence level (0.05 is a standard choice) is suitable for large data sets, this procedure must be treated with care when the data samples are small and difficult to obtain \citep{Null}.  A well-known way for estimation of parameters is to use the method of moments \citep{CaseBerg:01}, which yields potentially biased but consistent estimators.  Another method is bootstrapping from existing data to estimate the parameters of the parent distribution via the plug-in principle \citep{plug-in}.  However, these are often of little help for determining the amount of data needed to gain a specified level of significance, nor which targets we should favor when collecting this data.  Modelling the output of future missions can help to proactively determine the amount needed, and the authors regard this as a necessary step of the mission planning process, but this is laborious, must be redone from scratch for each new mission, and, in our opinions, does not always lead to an enlightening understanding of why a certain amount of data is required.  Our framework aims to augment these existing procedures and guide intuition by showing how back of the envelope calculations can lead us to concrete, actionable recommendations for mission design.

We now demonstrate the wider applicability of our framework by applying it to extensions of these simple setups.

\section{Ratio of 2 Signals}\label{section_ratio}

As a first application of our framework, we turn our attention to the case where we want to determine how the occurrence rate of a biosignature depends on some system parameter.  This is a generalisation of the analysis done in \citep{Sandora_2020}, which made the idealisation that all systems being measured were equally probable of hosting life.  Here, we are interested in determining the minimum number of systems required to be observed to detect a population trend, and how best to allocate mission time toward systems of varying characteristics to most effectively distinguish whether a trend exists.

This analysis has many applications.  Indeed, we may expect the probability of biosignatures to depend on planet mass, stellar mass, incident irradiation, planet temperature, age, metallicity, etc.  For illustrative purposes, we focus our attention on a particular case, the system’s position within our galaxy, though the analysis will be general enough to easily adapt to any of the other quantities mentioned.

There are numerous schools of thought on how habitability may depend on galactic radius (for a review, see \citep{habitability_net}).  Metallicity (the abundance of elements higher than hydrogen) is observed to decrease with galactic radius \citep{daflon2004galactic}, which may either be a boon or bane for life.  On the one hand, a minimum amount of material is required for planet formation \citep{Johnson_2012}, and planets are expected to be more abundant around more metal-rich stars \citep{life10080132}.  On the other hand, the prevalence of hot Jupiters also has been shown to increase with metallicity and, depending on the formation mechanism, could preclude the presence of terrestrial planets in the habitable zone \citep{fischer2005planet}.  Additionally, galactic centres (usually) host an active galactic nucleus (AGN), which spews high energy radiation that can strip planetary atmospheres \citep{habitability_net}.  These two competing factors have lead to the hypothesis that there is a galactic habitable zone \citep{GHZ}, analogous to the circumstellar habitable zone, comprised of a possibly thin annular region, only within which habitable planets can exist.  This hypothesis has drawn some criticism, not least because stars are known to migrate radially throughout the course of their evolution, defying simple attempts at placing their orbit \citep{prantzos2008galactic}.

Here, we treat the GHZ as a hypothesis, and ask what sort of data we would need to either verify or falsify it.  We may divide the alternatives into four scenarios: (i) null hypothesis- the biosignature occurrence rate does not depend on galactic radius (ii) Z- the occurrence rate is greater toward the center (iii) HJ- the occurrence rate is greater toward the edge, and (iv) AGN- the occurrence rate is 0 inside some critical radius.

A fully statistically rigorous approach would do something along the lines of fitting a curve for occurrence rate versus galactic radius, then determining to what degree of certainty the slope is away from 0.  Here, we opt for a much simpler method of bucketing our population into two bins, then comparing the occurrence rates of each bin and determining if they are different from each other to the statistical power afforded by the mission.  While this simplified analysis is not a replacement for the full approach, it allows us to easily track the dependence on mission parameters, and so can be useful for planning purposes.

We therefore assume we have two populations, either from the same or different missions.  If mission 1 surveys $\mtot$ planets and detects $\Mdet$ signals, and mission 2 surveys $\ntot$ planets and detects $\ndet$ signals, the signal occurrence rate in each population will be given by Beta distributions $X_M\sim\beta(\mtot,\Mdet,f)$ and $X_N\sim\beta(\ntot,\ndet,f)$ respectively.  Then the inferred ratio of these two occurrence rates $X=X_M/X_N$, is given by the following distribution \citep{doi:10.1080/03610920008832632}:
\beq
    X\sim
\begin{cases}
    \frac{B(\alpha,\beta_N)\,{ } _2F_1(\alpha,1-\beta_M;\alpha+\beta_N;x)}{B(\alpha_M,\beta_M)B(\alpha_N,\beta_N)}x^{\alpha_M-1},& x\leq 1\\
    \\
    \frac{B(\alpha,\beta_M)\,{ } _2F_1\left(\alpha,1-\beta_N;\alpha+\beta_M;\frac{1}{x}\right)}{B(\alpha_M,\beta_M)B(\alpha_N,\beta_N)}x^{-\alpha_N-1}, & x>1
\end{cases}
\eeq
where $\alpha_M=\Mdet+1$, $\beta_M=\mtot-\Mdet+1$, $\alpha_N=\ndet+1$, $\beta_N=\ntot-\ndet+1$, $\alpha=\alpha_M+\alpha_N$.  The hypergeometric function is $_2F_1(a,b;c;x)=\sum_{n=0}^\infty(a)_n(b)_n/(c)_nx^n/n!$, with $(a)_k=\Gamma(a+k)/\Gamma(a)$ the k-th order rising Pochhammer symbol.

The moments of this distribution can be written as
\beq
E\left[X^k\right] = \frac{(\Mdet+1)_k}{(\mtot+2)_k}\frac{(\ntot+2-k)_k}{(\ndet+1-k)_k}
\eeq

From this we can derive the mean
\beq
\mu=\frac{\Mdet+1}{\mtot+2}\frac{\ntot+1}{\ndet}=\frac{f_M}{f_N}+\mathcal O(1/N)
\eeq
where $f_M = M_\tu{det}/M_\tu{tot}$ and $f_N=N_\tu{det}/N_\tu{tot}$

Similarly, the variance can be written
\bea
\sigma^2&=& \frac{(\Mdet+1)(\Mdet+2)\ntot(\ntot+1)}{(\mtot+2)(\mtot+3)(\ndet-1)\ndet}-\mu^2\nonumber\\
&=&\frac{f_M^2}{f_N^2}\left(\frac{1-f_N}{f_N\ntot}+\frac{1-f_M}{f_M\mtot}\right)+\mathcal O\left(1/N^2\right)\label{ratvar}
\eea

The variance is not symmetric about the two ratios $f_M$ and $f_N$ since the associated distribution $X_M/X_N$ is not symmetric.  However, the quantity $\sigma^2/\mu^2$ is symmetric, and has a characteristic 1/N behaviour, but in this case dependent on the number of detected signals $\ndet=f_N\ntot$ and $\Mdet=f_M\mtot$, rather than the total number of systems observed.  Note also that this variance contains two additive contributions from the uncertainties in both surveys.

If one occurrence rate is smaller than the other, a correspondingly larger number of total systems need to be surveyed to adequately measure any trend with system parameters, rather than the overall occurrence rate.  For rare signals, $f_{M,N}\ll1$, the number of detections in each survey should be roughly equal.  This implies that surveys should collect more signals from regions where the signal is expected to be rarer.

To use this to distinguish between two competing hypotheses predicting different values of the ratio $f_M/f_N$, we would want the measurement error $e=\sqrt{\sigma^2}$ to be smaller than the difference in predictions.  So if we are trying to distinguish between the different galactic habitability scenarios outlined above, we would want $\sigma\lesssim\mu$.

\subsection{Cost analysis}

\begin{figure*}
    \centering
    \includegraphics[width=\textwidth]{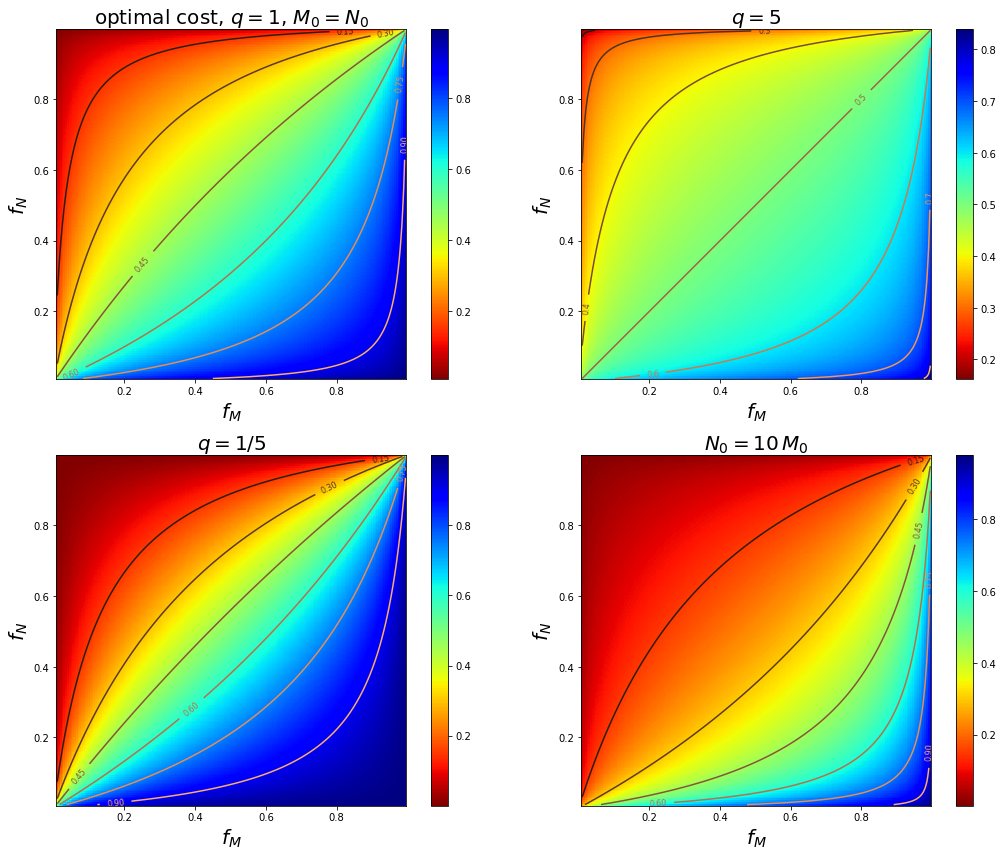}
    \caption{Optimum cost fraction $c$ to minimise variance. Unless specified, these plots take $q=1$ and $M_0=N_0$. Here, a low value of $c$ corresponds to allocating most funds to mission $M$.}
    \label{costplot}
\end{figure*}

We now use this analysis to determine the optimal way of apportioning funds to the two populations.  We denote the fraction of funding given to mission 2 as $c$, and as an example assume the cost functions for these surveys are given by the same power law relation $M=M_0(1-c)^q$ and $N=N_0c^q$, where $M_0$ and $N_0$ are the yields that would be obtained if all resources were put into mission $M$ and $N$, respectively.  As we discuss in section \ref{money}, we expect $q>1$ if the two populations come from two different missions, as increasing budget allows multiple mission aspects to be improved simultaneously.  For apportioning observing time within the same mission, we instead expect $q<1$, as increasing observing time yields diminishing returns, discussed further in \citep{Sandora_2020}.  We wish to minimise the variance in equation (\ref{ratvar}), corresponding to maximum information gain.  The derivative of eqn (\ref{ratvar}) yields
\bea
     \frac{\partial}{\partial c }\sigma^2
    &=& \frac{f_M^2}{f_N^2}\bigg[\frac{-q}{N_0}\frac{1-f_N}{f_N}c^{-(q+1)}\nonumber\\
    &&+\frac{q}{M_0}\frac{1-f_M}{f_M}(1-c)^{-(q+1)}\bigg]
\eea
This makes use of the approximation that both $\Mdet$ and $\ndet$ are substantially removed from 0.  Setting this to zero and solving for $c$ gives us the optimal cost
\beq
c=\frac{1}{1+s^\frac{1}{q+1}},\quad s=\frac{N_0}{M_0}\frac{f_N}{f_M}\frac{1-f_M}{1-f_N}
\eeq
Yielding survey sizes
\beq
N = N_0\frac{1}{\left(1+s^{\frac{1}{q+1}}\right)^q},\quad M = M_0\frac{s^{\frac{q}{q+1}}}{\left(1+s^{\frac{1}{q+1}}\right)^q}
\eeq
From here, it can be seen that if $f_M=f_N$ and $\mtot=\ntot$, then $c=1/2$, i.e. both surveys should get equal priority.  The relative cost is plotted for generic values of observed fraction in Fig. \ref{costplot}.

This allows us to explore how the optimal cost depends on various parameters.  The recommendations are sensible - the better the observations in population M, the more cost should be invested on N and vice versa to get maximum information.  If $q$ is large, the optimal cost will deviate more strongly from 1/2 compared to the $q=1$ case, holding other parameters fixed.  Similarly if $q$ is small, we have $c\approx1/2$ over a broad range of parameter values.  In the lower right panel we see that more money should be spent on missions that have lower yields.

Lastly, note that from the above expression, we see that the optimal survey sizes depend on biosignature occurrence rates, which are not necessarily known beforehand.  If these are unknown, a well designed survey would be able to adapt to information as it arrives, to alter upcoming target selection as current estimates of the occurrence rates dictate.

\section{The Tradeoff Between False Positives and False Negatives}\label{section_tradeoff}
In this section, we apply our procedure to solar system planetary biosignature missions.  Here, we address ambiguity in the interpretation of mission results. This ambiguity can arise from two significant contributing factors. Firstly, there could be biosignature experiments where a positive result may be a false positive, \textit{e.g.} some geochemical trickery masquerading as life. Alternatively, there may be situations where a negative result could be a false negative, \textit{e.g.} there may be entire communities of microbes present which the instrument is not sensitive enough to detect. Any realistic mission will have to deal with both.  In such a scenario, and assuming finite resources/design constraints, one key decision which will have to be made is which of the two sources of ambiguity to prioritise minimising to arrive at a higher overall certainty of whether or not life is present. As discussed in \citep{foote2022false}, the need to mitigate these leads to the drive toward the most definitive biosignatures possible, and as exhaustive a determination of abiogenesis rate as possible.

Unlike large scale surveys of many exoplanet systems, solar system based biosignature search strategies are characterised by information depth rather than information breadth. In  other words, in the solar system life detection mission, we are going to be successively surveying the same planet over and over again, gaining more and more information and constraining our statistical parameters. For this reason, the ideal framework from which to approach the problem is a Bayesian one, which can successively adjust prior probabilities based on new information \citep{catlingdavid2018exoplanet}.

We define the false positive rate as the probability of a detection given nonlife, $\FP=P(D|NL)$ and the false negative rate as the probability of a nondetection given life, $\FN=P(ND|L)$. The quantities needed for the information can be derived via Bayes' theorem:
\beq
P(L \mid D,FP,FN,f)=\frac{P(D \mid L,FP,FN,f)P(L \mid FP,FN,f)}{P(D \mid FP,FN,f)}
\eeq
Where
\beq
P(L \mid FP,FN,f) = P(L) = f
\eeq
is the probability of life existing in the location where the probe searched.  The probability that the instrument registers a signal is
\beq
P(D \mid FP,FN,f) = (1-\FN)f + \FP(1-f)
\eeq
Therefore, the posterior probability of life given a detection is given by
\beq
P(L \mid D,FP,FN,f) = \frac{(1-\FN)f}{(1-\FN)f + \FP(1-f)}
\eeq
and
\beq
P(L|ND,FP,FN,f) = \frac{\FN f}{\FN f + (1-\FP)(1-f)}
\eeq

To arrive at a decision making framework, we approach the problem in terms of information gain. Here, the information gain measures the uncertainty present in data–the larger the Shannon Entropy, the lower the certainty of the data and the higher the likelihood that it may be random.  In this case the probabilities are conditional, so we use the Shannon Entropy for conditional probabilities, given in terms of the event of detection, nondetection, and the presence of life below.  This is the information of the posterior probability distributions, weighted by their probability of occurrence.
\bea
S(\FP,\FN) = -\int df p_f(f) \bigg[P(D)\,s\big(P(L|D)\big) \nonumber\\
+ P(ND)\,s\big(P(L|ND)\big)\bigg]
\eea
where $s(p)=p\log p +(1-p)\log(1-p)$.

\begin{figure*}
\begin{center}
\includegraphics[width=1.0\textwidth]{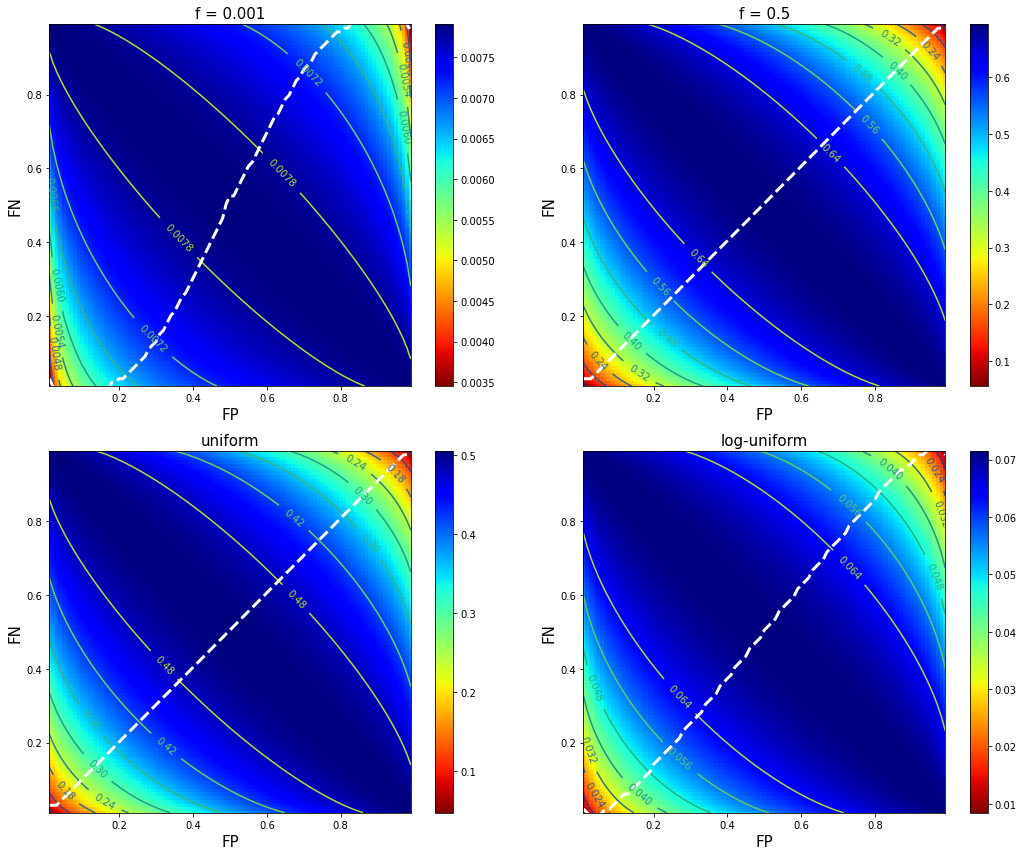}
\caption{Heat map of the conditional entropy in terms of FP and FN.  The top two plots take particular values for the probability of life existing at the sampling location $f$.  The bottom two integrate over a uniform and log-uniform distribution for $f$, respectively.  The white dashed line in each separates regions where it is more beneficial to decrease FP (above the line) and FN (below the line).}
\label{conde}
\end{center}
\end{figure*}

Lastly, as the probability of life $f$ is not known ahead of time, our expression requires that we integrate over all possible values.  This requires a prior distribution $p_f(f)$, which incorporates our current estimates on the probability of life existing in a particular locale, given the current uncertainties in the probability of life emerging, as well as any geological history which may have impacted the site's habitability \citep{westall2015biosignatures,lorenz2019calculating}.  This is an inescapable part of the calculation, and so to investigate the dependence of our results on any assumptions we make about this quantity we explore four different options: (i) the value of $f$ is arbitrarily taken to be .001 (ii) the value of $f$ is taken to be .5 (iii) the probability of any value of $f$ is uniformly distributed, and (iv) the probability is log-uniform, $p_f(f)\sim1/f$.  This last option requires a minimum value for $f$, which we take to be $10^{-10}$ in our calculations.  The choice of this value is arbitrary, but mainly influences the overall scale of the expected information gain through $S\sim1/\log(p_\text{min})$, rather than the dependence on the variables FN and FP.  The expected information gain for these three choices is displayed in Fig. \ref{conde}.

In these plots, the blue regions throughout the midsection have highest entropy, corresponding to maximum ambiguity in result interpretation. The red regions in the upper right and lower left corners are both places where the information entropy approaches zero, corresponding to certain result interpretation. The rightmost red region is a pathological point in parameter space where both FP or FN approach 1. In  our parameterization, this region indicates that the signal (or lack thereof) is almost sure to convey the opposite of what the instrument was designed to do, and the experiment is certain to be wrong. This is what every biosignature mission should deliberately avoid.

In the bottom left corner, we approach our second region where $S\rightarrow0$, this time where FP and FN both approach zero, representing the ideal scenario. Such unambiguous results will be unobtainable for any realistic mission, but these plots do provide a metric that can be used to evaluate missions with different values of FP and FN. For a preexisting mission design that can be improved upon incrementally, the recommendation given by this analysis is to go in the direction of maximum gradient, which depends on the mission's position in the plot. Through these plots, we have placed a white dashed line demarcating the region where it is more important to prioritise false positives (above the curve) from the region where it is more important to prioritise false negatives (below).  This line is exactly equal to the diagonal FP$=$FN for the uniform and $f=.5$ cases, but is substantially different in the others.  Notably, in the region where both FP and FN are small, it remains more important to prioritise false positives.  This is a consequence of our assumption that $f\ll1$; the opposite conclusion would hold if we consider $f\approx1$.

This procedure, is, of course, reliant on a method of estimating false positive and negative rates for a mission, both of which may be difficult in practice.  FP can be estimated by characterising the instrumental sensitivity, and accurately accounting for all abiotic signal sources with careful geochemical modelling.  This is often fraught with controversy (for a recent review see \citep{harman2018biosignature}), as historical examples can attest (see e.g. \citep{barber2002origin,yung2018methane,bains2021astrobiologists}).  FN, on the other hand, may be just as difficult to ascertain, as life may be present in far lower abundances than Earth based analogue environments indicate, yielding uncertainty in overall signal strength \citep{cockell2009cryptic}.  Additionally, extraterrestrial life may have entirely different metabolisms, and so may not produce a targeted biosignature \citep{johnson2019agnostic}.  

As further illustration of the application of this formalism, we next turn to two more examples: the combination of multiple instruments, and the recent detection of methane on Enceladus.

\subsection{Combining Signals from Multiple Instruments}

Here we investigate the effect of combining signals to mitigate false positives and negatives, which can provide redundancy and ameliorate the difficulties in interpretation present for a single signal in isolation.  Generally, this can be done in a variety of different ways, to either increase the coverage or certainty of a detection.  We illustrate the various approaches in a simple setup, and show that our framework yields recommendations for which choice is preferable.

Let's restrict our attention to the combination of two signals.  The first is fixed, say the detection of a certain biosignature (say, methane), which carries with it a probability of false positive and false negative.  We then have two options for including a second instrument.  In option A, we're concerned with false positives, so we include a second instrument to detect a secondary signal which we expect to be present in the case of life (say, homochirality).  We then claim detection of life only if both these signals register.  While this increases our confidence in a detection, it comes at an increased risk of false negatives, as if either instrument fails to register, we would reject the signal of the other.

Alternatively, we may be more concerned with false negatives, and so may include an additional instrument to measure a third biosignature (say, a spectral red edge), which would be complimentary to the first.  This improves our original coverage, and so alleviates concern for false negatives.  However, this also increases the chances of false positives, for if either instrument registers a signal, we would claim detection.

The question we wish to address is which of these two options is preferable, given the parameters of the setup.  For simplicity, we assume that the false positive and false negative rates of all three instruments are identical, to avoid a proliferation of variables.  We also assume that the signals which they measure are uncorrelated with each other (as a high degree of correlation would render additional instruments not worthwhile).  In this case, we have
\bea
P_A(\FP) = \FP^2,\quad P_A(\FN)=1-(1-\FN)^2\nonumber\\
P_B(\FP) = 1-(1-\FP)^2,\quad P_B(\FN)=\FN^2
\eea

\begin{figure}
\begin{center}
\includegraphics[width=.5\textwidth]{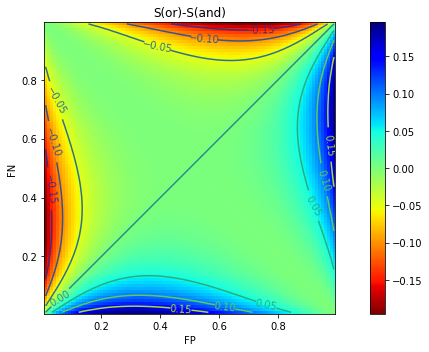}
\caption{Difference in two signal entropy for two different signal combination strategies.  If this difference is positive, it is more beneficial to use the ``and'' strategy, so that both signals must register to count as a detection.  If the difference is negative, the ``or'' strategy, where either signal registering would count as a detection, is preferable.}
\label{Sdiff}
\end{center}
\end{figure}

With this, we display the difference in information gained from these two setups in Fig. \ref{Sdiff}, using the value $f=.001$.  We notice that if $\FP>\FN$, then scenario $A$ is preferred, where both signals must register to count as a detection.  Also of note is that the difference in the two setups is greatest when either one of the two probabilities is either very small or very large.

\subsection{Direct Application: The Enceladus Data}

The Cassini spacecraft, as part of its mission to study Saturn and its moons, passed through a plume erupting from the south pole of Enceladus to collect a sample. Instrumental analysis of the sample revealed a relatively high abundance of methane (CH4), molecular hydrogen (H2) and carbon dioxide (CO2) \citep{waite2017cassini}. Methane is considered a potential biosignature because on Earth, microbial communities around hydrothermal vents obtain energy through chemical disequilibrium present in H2 emanating from the vents, resulting in the release of methane as a waste product, a process known as methanogenesis \citep{lyu2018methanogenesis}. However, this is not an unambiguous signal, as methane can also be produced abiotically in hydrothermal vent systems via serpentization \citep{article}.  One method of distinguishing between these two scenarios is then to compare the ratio of fluxes, as biogenic production would yield a higher relative methane rate than would occur abiotically.

This situation represents an ideal test case for our decision making tool--an ambiguous biosignature result, with potential for both false positives and false negatives in the data.

An analysis of the Cassini findings by \citep{affholder2021bayesian} compared outgassing rates of simulated populations of archaea and abiotic serpentization, and claimed moderate support for the biotic hypothesis on Bayesian grounds.  The model also indicated that at the population sizes and rates of methanogenesis demanded by the data, the archaea would have a negligible effect on the abundance of H2--therefore, the high concentration of H2 present in the data did not count against the biotic methanogenesis hypothesis.

It should be noted that this model still carries considerable ambiguity, and a substantial portion of their model runs produce the observed signal abiotically.  Regardless, they present quantitative estimates for the probabilities of false positives and false negatives in actual data. Therefore, it is well suited for an application of our decision making tool.

\begin{figure}
\begin{center}
\includegraphics[width=.5\textwidth]{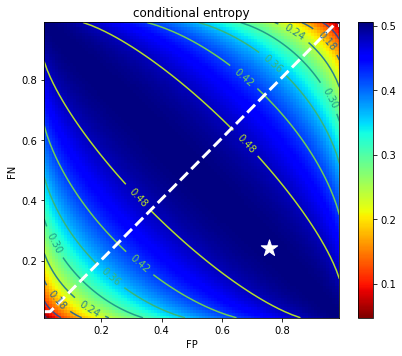}
\caption{White star corresponds to the estimated FP and FN of the Enceladus Cassini flythrough mission.  The underlying plot is the same as shown in Fig. \ref{conde}, with uniform prior for $f$.}
\label{encel}
\end{center}
\end{figure}

According to the biogeochemical models in \citep{affholder2021bayesian}, the Cassini plume flybys resulted in FP=.76 and FN=.24.  Here FP and FN are determined by the limits of instrumental sensitivity; from \citep{waite2017cassini} the sensitivity threshold is estimated to be about 10x smaller than the observed values.  This is illustrated in Fig. \ref{encel}, where we place a white star to indicate where in the diagram out current knowledge of these biosignatures lies.

\section{Lurkers}\label{section_lurkers}

As a further application, we apply our approach to the spatial problem, where we are concerned with the tradeoff between coverage and resolution when looking for a biosignature.  Much of the language we use in this section is couched in terms of technosignature ``lurkers'', but our formalism is general and can equally be applied to microbial colonies, molecular biomarkers, etc.

\subsection{Background and Context}
An ongoing area of technosignature research considers physical artefacts and probes which may be hiding in our solar system, referred to as ``lurkers'' \citep{benford2019looking}. The idea is that if extraterrestrial civilisations are present in our galaxy, they may have seeded the galaxy with probes, which, either defunct, dormant or active, could be present in our solar system \citep{haqq2012likelihood}. The plausibility of this was argued by Hart and Tippler, for the time it takes self replicating von Neumann devices to cover the galaxy is modelled to be much shorter than the age of the galaxy itself \citep{gray2015fermi}. Therefore, in spite of the considerable energy requirements and engineering challenges inherent in interstellar travel, a reasonable case can be made for past extraterrestrial visitation of our solar system which may have left behind artefacts \citep{freitas1983search}. Hart and Tippler's argument frames the absence of such lurkers as justification for the Fermi Paradox and pessimism about the prevalence of intelligent life in general \citep{gray2015fermi}. However, as deduced by \citep{haqq2012likelihood}, only a small fraction of the solar system's total volume has been imaged at sufficient resolution to rule out the presence of lurkers. Therefore it remains an open question warranting further investigation and dedicated search programs.
 
This motivates us to apply our formalism to this setup to determine how to effectively allocate resources in order to maximise our chances of finding signs of lurkers. While a survey with unlimited resources may be able to scour every square inch of a planet (or moon, asteroid, etc.), any realistic mission will have budgetary and lifetime constraints.  The two extreme strategies would then be to perform a cursory inspection across the entire planet area, but potentially miss signals with insufficient strength, or to perform a high resolution search in a small region of the planet.  Here we show that the ideal mission forms a compromise between these two extremes, and deduce the optimal design to maximise chances of success (which in our language corresponds to the mission yielding most information).
  
\subsection{Mathematical Framework}

The setup we consider consists of a survey of resolution $R$, which is the two dimensional spatial resolution of a survey.  Our survey covers a fraction of the planet area $f_R$.  We assume that our survey results in no detection (as otherwise, our concern with interpretive statistics becomes lessened considerably).  We would like to compute the probability that at least one object is present, given a null result, $p(N>0|\text{null})$, and want to optimise mission parameters so that this quantity is as small as possible given some constraints.  Of course, this is equivalent to the formulation where we try to maximise the quantity $p(N=0|\text{null})$.

To compute this, first note that, in the scenario where there are $N$ objects uniformly spaced on the planet surface, all of size $D$, we have
\beq
p(\text{null}|N,D) = (1-f_R)^N \theta(D-R)+1\,\theta(R-D)
\eeq
Here $\theta(x)$ is the Heaviside function, which is 1 for positive values and 0 for negative values.  From here, we can see that a null result is guaranteed if our mission resolution is above the object size.  If the mission does have sufficient resolution, we see that the probability of a null signal diminishes with increasing survey coverage or number of objects.

Inverting this expression requires priors for both the number of objects and their size.  If the prior distribution of object sizes is $p_D(D)$, this expression can be integrated to find
\bea
p(\text{null}|N)&=&\int dD\, p_D(D)\, p(\text{null}|N,D)\nonumber\\
&=& 1-\left(1-(1-f_R)^N\right)P_D(D>R)
\eea
And if the prior distribution of number of objects is $p_N(N)$, the total probability of nondetection is 
\beq
p(\text{null}) = \sum p(\text{null}|N)\,p_N(N).
\eeq
Then by Bayes' theorem,
\beq
p(N|\text{null})=\frac{p(\text{null}|N)p(N)}{p(\text{null})}.
\eeq
Note that here, in the limit $R\rightarrow\infty$, the survey is completely uninformative, and we have $p(N|\text{null})\rightarrow p(N)$.

Then, the probability of at least one object being present given a null result is
\beq
p(N>0|\text{null})=1-\frac{p_N(N=0)}{p(\text{null})}.
\eeq

In the limit of complete survey $f_R\rightarrow1$, we find
\beq
p(N>0|\text{null})\rightarrow\frac{P_D(D<R)}{\Theta+P_D(D<R)},\quad \Theta=\frac{P_N(N=0)}{P_N(N>0)}
\eeq
reproducing the results of \citep{haqq2012likelihood}.

To simplify this general analysis, we now specialise to the case where the number of objects is constrained to be either 0 or $N^*$, $p_N(N)=(1-p^*)\delta_{N,0}+p^*\delta_{N,N^*}$, where $\delta_{i,j}$ is the Kronecker delta.  Then we have
\beq
p(N>0|\text{null})=\frac{p^*\,\left(P_D(D<R)+(1-f_R)^{N^*}\,P_D(D>R)\right)}{1-p^*\,\left(1-(1-f_R)^{N^*}\right)\,P_D(D>R)}\label{plurk}
\eeq

\begin{figure*}
\begin{center}
\includegraphics[width=\textwidth]{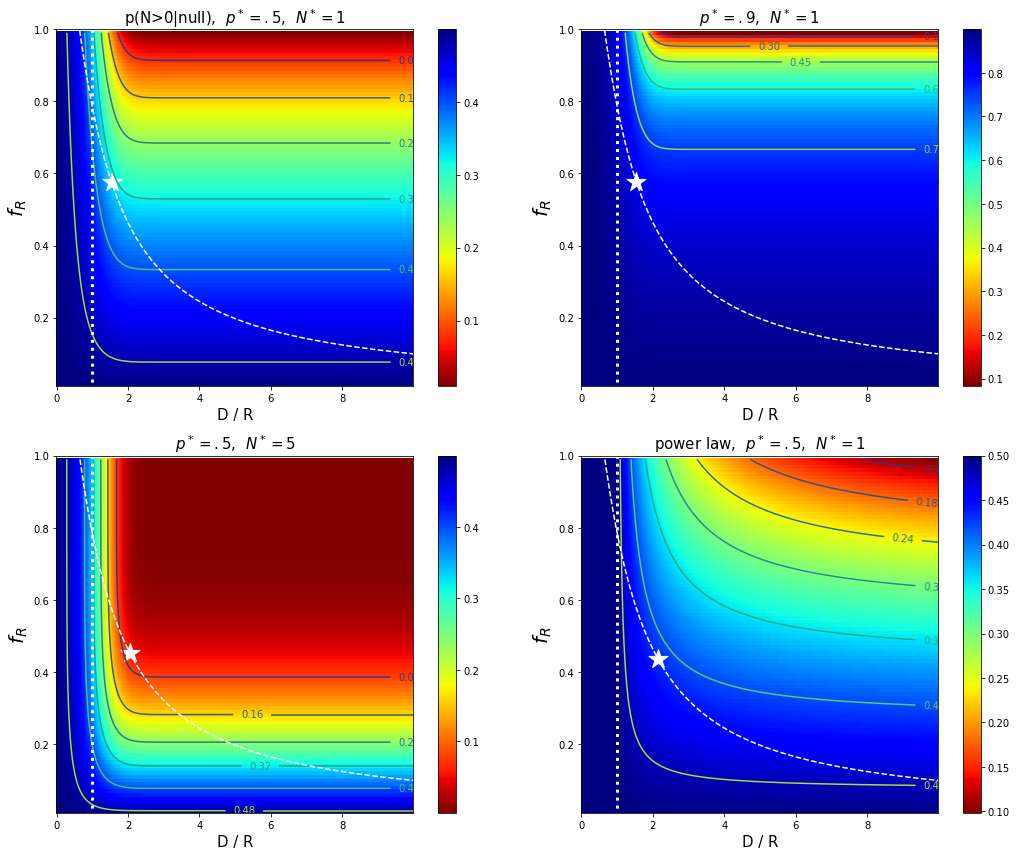}
\caption{The probability that lurkers exist on a surface as a function of mean diameter over resolution $D/R$ and fraction of surface covered $f_R$.  The vertical dashed line corresponds to mean lurker diameter, and the dashed curve corresponds to a constraint on mission design.  The white star is the point of lowest $p(N>0|\text{null})$ restricting to the constraint.  The top two and bottom left plots take a normal distribution for lurker diameter, and the lower right plot takes a power law.}
\label{lurker_plots}
\end{center}
\end{figure*}

We display this expression for various parameter values in Fig. \ref{lurker_plots}, which plots the ratio of mean lurker size to resolution $D/R$ versus the coverage fraction $f_R$.  In the first three panes, we assume the distribution of lurker diameters follows a normal distribution (with mean 1 and standard deviation .5).  The aim of a mission should be to get as close to the `red region' of these plots as possible given mission constraints.  We overlay a constraint curve $A\sim \tan^{-1}(1/R)$ to encapsulate a typical area/resolution tradeoff, but this form specifies a family of curves with differing base values.  For a given constraint curve, this will single out a unique mission design that minimises this quantity (thus maximising science return).  This point is denoted by the white star in each of these plots.

Comparing the top two plots, we can see that the optimum point does not depend on the lurker probability $p^*$ at all, a fact that can be derived from the form of eqn. (\ref{plurk}).  Comparing to the lower left plot, we can see that the optimum point is shifted toward lower $f_R$ and higher $D/R$ for larger $N^*$, which is reasonable, as there would be a greater probability of spotting a lurker in a given area if more are present on the planet surface.  The bottom right plot assumes a power law distribution (with $p(D)\propto D^{-2}$) of lurker sizes to highlight the effect this has on our recommendation; again, this shifts the optimum point toward lower $f_R$ and higher $D/R$. 

\subsection{Direct Application to Solar System Bodies}

Here, we will demonstrate a direct application of this statistical framework, by examining the recommendations we find for a series of different bodies in the solar system. We use the spatial resolutions of planetary missions and surveys to establish upper limits on the probability of lurkers which may have gone undetected up until now, allowing us to make a series of recommendations for each solar system body/region of space.  For this, we display the resolution and fraction of area covered for multiple missions to bodies around our solar system in Table \ref{value_table}.

\begin{table*}
	\centering
        \caption{Upper limit diameters for lurkers on the surfaces of solar system bodies and regions of surveyed space.  Original concept for this table and all numbers adapted from \citep{lazio2022techno}.}
        \begin{tabular}{|c|c|c|c|c|}
	\hline
	Location & R & $f_R$ & Mission & Source\\
	\hline
	Mercury & 250 m &1 & MESSENGER & \citep{0973d9680f4e421b993d3fe06b10f659}\\
 	Venus & 10 m & $10^{-9}$ & Venera &\citep{freitas1985there}\\
          & 250 m & 1 & Magellan & \citep{saunders1992magellan}\\
	Earth & 1 cm & 1/3 & various &\citep{freitas1985there}\\
	Moon & 0.5 m & 1  & LRO & \citep{robinson2010lunar}\\ 
        Mars & 150 m &.86 & MRO & \citep{zurek2007overview}\\
		 & 1 m &.01 & HiRISE &\citep{kirk2008ultrahigh}\\
	Ceres & 35 m & .97 & Dawn &\citep{park2019high}\\
        Europa (Jupiter) & 230 m & .01 & Galileo &\citep{figueredo2000geologic}\\
	Enceladus (Saturn) & 110 m  & .96 & Cassini &\citep{roatsch2008high}\\
	Ganymede (Uranus) & 3.5 km & .94 & Galileo & \citep{patterson2010global}\\
        Triton (Neptune) & 3 km & 1/3 & Voyager 2 &\citep{abelson1989voyager}\\
        Pluto & 660 m & .45 & New Horizons &\citep{olkin2017global}\\
         & 77 m & .003 & New Horizons &\citep{keeter2016}\\
        \hline
	\end{tabular}
	\label{value_table}
\end{table*}

The estimates quoted are for illustrative purposes only, and there are some nuances which the table does not account for. Our model makes a simplifying assumption by assuming that lurkers will be unambiguously distinguishable from background geological features.  This may not necessarily be true, especially if deliberately camouflaged. For the same reason, even excluding the possibility of deliberate camouflage or stealth, using diameter as our only proxy for detectability has its limitations because the lurkers' albedo relative to that of its surroundings is also a factor. A low albedo lurker on a dark planetary surface may blend in, appearing to be a shadow of another feature or a dark rock. Conversely, a lurker with an albedo that contrasts more with surrounding geology may be easier to detect. Additionally, these are remote sensing missions, and so any subsurface/sub-ocean area would be inaccessible and are not accounted for in this application.

\begin{figure*}
\begin{center}
\includegraphics[width=\textwidth]{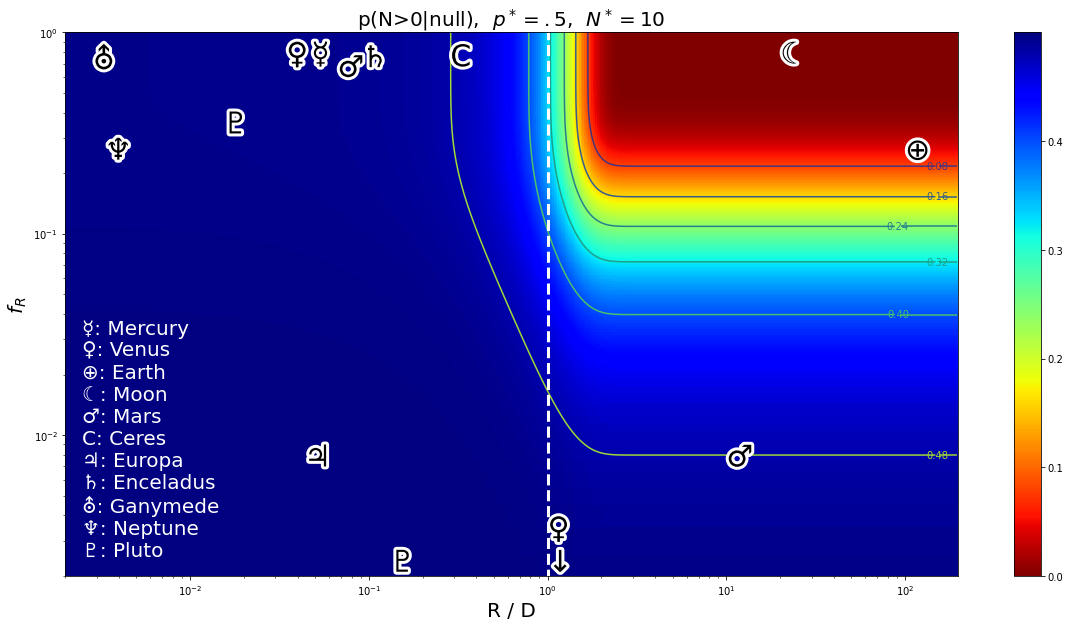}
\caption{Probability that lurkers exist on the various solar system bodies listed in Table \ref{value_table}.  This assumes an a priori probability of 1/2, and considers the scenario where 10 lurkers would be present, with a mean diameter of 10 meters.}
\label{lurker_points}
\end{center}
\end{figure*}

The values in Table \ref{value_table} are displayed alongside our estimate for $p(N>0|\text{null})$ in Fig. \ref{lurker_points}.  As fiducial values, we use $p^*=.5$, $N^*=10$, and $\langle D\rangle=10$ m, which is a reasonable upper estimate for the size of an interstellar probe \citep{freitas1983search}.  From this, we can draw a number of conclusions:

\begin{itemize}
\item Of the bodies considered, we are most certain that no lurkers exist on the moon, followed by Earth.  This is because although the Earth has been mapped to higher resolution on the continents, the oceans prevent coverage for the majority of the surface.
\item The gradient at each point dictates which aspect of a particular mission should be improved.  In a nutshell, whichever is worse between the covered fraction and resolution is what should be improved, but this formalism allows for that comparison to be made quantitatively.
\item Several bodies have near-complete low resolution maps, with small patches of high resolution.  This allows us to compare which are more valuable for confirming the absence of lurkers.  For Mars, for example, the high resolution images give us most confidence that lurkers are absent.
\item Of the bodies considered, we are least certain about the absence of lurkers on the outer solar system moons. Areas even further out, such as the Kuiper belt and Oort cloud, are even less certain.
\item Despite mapping efforts, many of these bodies still have $p(N>0|\text{null})\approx.5$, which is our reference probability that lurkers are present.  Therefore we are nowhere close to being able to claim that no lurkers are present in the solar system.
\end{itemize}

We may concern ourselves with the overall probability that lurkers are present in the total solar system, which is given by the compound probability of each body separately, $p(N>0|\text{null})_\text{tot}=1-\prod_{b\in\text{bodies}} (1-p(N>0|\text{null})_b)$.  This total probability is most sensitive to the body with lowest probability, suggesting that if our goal is to minimise this quantity, we ought to prioritise the bodies we are least certain about first.

\section{Further Applications}\label{section_further}

\subsection{What is the best way to determine a habitability boundary?}\label{boundary}
	
	Supposing we do begin to detect biosignatures, we will become interested in the problem of determining the range of conditions for which life can persist.  Though there are a number of variables that certainly play a role (temperature, planet size, star size, atmospheric mass, water content, ellipticity, obliquity, etc.), in this section we focus on a single abstract quantity $t$.  We wish to outline a procedure for determining the range of $t$ which can sustain life.  For simplicity we focus on determining the lower bound, denote the current range as $(0,t_1)$, and treat the variable as uniformly distributed\footnote{if a quantity is not uniformly distributed, its cumulative distribution function can be used as $t$ everywhere in this section.}.
	
	In the scenario where all systems with habitable characteristics are guaranteed to possess biosignatures, this problem reduces to the standard method of bisecting the remaining interval until the desired precision is reached.  The subtlety here is that if a planet does not exhibit the biosignature, we would not be able to infer that life cannot exist within that range, because presumably life does not inhabit every potentially habitable environment.  Let us discuss the certain case first as a warm up, to outline how information gain can be used to arrive at this classic result.
	
	As stated, before our measurement we know that the lower bound for $t$ lies somewhere in the range $(0,t_1)$.  We will perform a measurement of a system with $t_0$.  A priori, the probability of a detection (the probability that the lower bound is less than $t_0$) is $P(\text{det})=t_0/t_1\equiv c_0$, and the probability of no detection is $P(\text{miss})=1-c_0$.  If a signal is detected, then the updated distribution of possible lower bounds is $P_\text{det}=\mathcal U(0,t_0)$, and if no signal is detected, the distribution becomes $P(t_\text{lower})=\mathcal U(t_0,t_1)$.  The expected information gain from this measurement is
	
	\bea
	\langle S\rangle &=& S(P_\text{det})P_\text{det}+S(P_\text{miss})P_\text{miss}\nonumber\\
	&=& -c_0\log c_0-\left(1-c_0\right)\log\left(1-c_0\right)
	\eea
	This quantity is maximised at $t_0=t_1/2$, recovering the classic bisection method.
	
	Now, let's generalise this to the case where life is only present on a fraction of all habitable planets, denoted $f$ (independent of $t$ in this section).  In this case, the probability of detection is given by $P_\text{det}=f c_0$, which is the product of the probability of the lower bound being smaller than the observed system, and that life is present.  The absence of a detection could be either because the lower bound is greater than the system's value, or because no life happens to be present.  So, the probability of no detection is given by the sum of two terms, $P_\text{miss}=P(t_0<t_l)+P(t_l<t_0)(1-f)=1-f c_0$.  The expected information gain is given by
	\bea
	\langle S\rangle = f c_0\log c_0+(1-c_0)\log(1-fc_0)\nonumber\\
 +(1-f)c_0\log\left(\frac{1-fc_0}{1-f}\right)
	\eea
	After some manipulation, it is found that this quantity is maximized at
	\beq
	c_0=\frac{(1-f)^{1/f-1}}{1+f(1-f)^{1/f-1}}
	\eeq
	When $f\rightarrow1$, this expression goes to $1/2$, reproducing the certain case.  When $f\rightarrow0$ it goes to $1/e$, reminiscent of the classic secretary problem, which aims to determine an optimal value in complete absence of information.
	
	The expected information gained by an optimal measurement is
	\beq
	\langle S\rangle|_\text{opt}=\log\left(1+f(1-f)^{1/f-1}\right) 
	\eeq
	This goes to $1/e$ for the certain case, representing 1 nat of information gain, and $f/e$ for small $f$, which is 1 nat multiplied by the detection probability.
	
	We can also determine the convergence rate toward the true lower bound by looking at the expected maximum possible value for a given time step.  Recall that in the certain case, this is given by $1/2t_0+1/2t_1=3/4t_1$, so that this converges as $(3/4)^N$ after $N$ time steps.  In the uncertain case, the formula instead yields $t_0 f c_0+t1(1-f c_0)$.  We do not display the full expression after substituting the optimal value of $c_0$ here, but instead report its limit for small $f$, which becomes $(1-(e-1)/e^2f)t_1$.  This series converges as approximately $\text{exp}(-.232fN)$, which is considerably slower than the certain case.

\subsection{What is the optimal amount of money to spend on different mission aspects?}\label{money}
	
	Another application is in determining how much money to spend on each aspect of a mission.  In this section we will abstractly define a number of mission aspects $\{a_i\}$, with the total number of observed planets scaling as $\ntot=N_0\prod_i a_i^{q_i}$.  Here, we assume that the total number scales with each aspect to some power.  For example, the aspects could correspond to telescope diameter $D_t$, mission lifetime $T$, spectral resolution $\Delta\lambda$, etc., as found in \citep{Sandora_2020} where $\ntot\propto D_t^{12/7}T^{3/7}\Delta\lambda^{3/7}$.  In the next subsection we will treat a step-wise increase as well.  Now, we assume that a mission has a fixed total budget $\ctot$ apportioned amongst each aspect as $\ctot =\sum_ic_i$.  We would like to determine the optimal way of assigning each $c_i$ so as to maximise the number of observed planets, as well as the resultant scaling of mission yield with cost.
	
	This is a straightforward convex optimisation problem, that can be solved inductively by finding the location where the gradient of the total yield vanishes for every component budget.  The solution is
	\beq
	c_i=\frac{q_i}{\qtot}\ctot,\quad 
	\ntot = N_0\frac{\prod_iq_i^{q_i}}{\qtot^{\qtot}}\,\ctot^{\qtot}
	\eeq
	Here, we have defined $\qtot=\sum_iq_i$, and assume nothing other than that each exponent is positive.
	
	This yields a concrete optimal fraction of budget to yield to each mission aspect.  Particularly noteworthy is the fact that the total yield scales superlinearly with budget for all practical applications of this formalism.  This implies, among other things, that the cost per planet decreases dramatically with increased scale.  The cost per information gain, however, does not, but is instead minimised at the value
	\beq
	c_\text{b}=\frac{e}{\left(\frac{\prod_iq_i^{q_i}}{\qtot^{\qtot}}\,N_0\right)^{1/\qtot}}
	\eeq
	This minimum occurs because the information return depends only logarithmically on total number, and so these diminishing returns must be balanced against initial startup costs.  For the reference scenario above, taking spectral resolution, telescope diameter and mission length (to each scale linearly with cost) gives $\qtot=18/7$, and that the optimal cost is $\$10^9$ when $N_0=121.8$.
	
	Several things to note about this expression: optimal pricing decreases slowly for missions with higher yield potential, and is smaller for missions that scale faster.  If followed, this gives the optimal number of planets for a mission to observe as $N_{opt}=e^{\qtot}$, which results in $\qtot$ nats of information.  However, minimising cost per information gain in this manner neglects the cost associated with mission development, which can be significantly higher than the costs for the duration of the mission.  This is included in section \ref{mvt}, where we find that the recommended number of systems to observe is substantially larger.

	\subsection{When should an additional instrument be included?}\label{included}
	
	In the previous subsection, we focused on mission aspects that affected the mission yield continuously and monotonically.  Some aspects do not scale this way; a particular case is which an aspect is either included or not, and its effectiveness cannot be improved other than it simply being present.  An example would be whether to include a starshade for a space mission, or whether to include an instrument capable of making a specific measurement that would lead to higher yields.
	
	First, we will consider the case that the type of biosignatures being measured are not affected by the inclusion of a binary aspect, but that only the total yield is.  In this case, we can say that the total yield is $\ntot(b,c) = N_0(b)c^{\qtot}$, where $b$ is the binary aspect, and $N_0(b)=N_0$ if the aspect is absent and $N_0(b)=N_1$ if it is present.  If the cost of the binary aspect is $c_b$, then this instrument should be included when
	\beq
	N_1\left(\ctot-c_b\right)^{\qtot}>N_0\,\ctot^{\qtot}
	\eeq
	
	It may also happen that a binary aspect does not affect the overall number of planets observed, but allows the measurement of a different class of biosignatures.  This is the case, for example, when deciding whether to include an instrument that can detect oxygen.  In this scenario, the yield without this instrument is given by $\ntot$, whereas with the instrument it is given by $\fbio\ntot^2$.  In this case, the instrument should be included when
	\beq
	\fbio\,N_0\,\left(\ctot-c_b\right)^{2\qtot}>\ctot^{\qtot}
	\eeq
	From this, we can see that the number of planets observed should be at least as large as $1/\fbio$, plus some to counteract the fact that fewer systems will be observed with the instrument.  If the occurrence rate is too low, the mission will not be expected to observe a signal anyway, and so the measurement of the second class of biosignatures would be a waste.
	
	\subsection{The marginal value theorem and mission lifetimes}\label{mvt}
	
	In this subsection we address the question of how long a mission should be run.  The yield scales sublinearly with mission lifetime, as the brightest, most promising stars are targeted first, leaving the stars which require longer integration times for later in the mission.  This gives diminishing returns as the mission goes on.
	
	The marginal value theorem from optimal foraging theory \citep{charnov1976optimal} can be applied to determine the point at which a mission should be abandoned in favour of a new mission with greater yield potential.  Indeed, collecting exoplanet data closely resembles the process of sparse patch foraging, wherein samples are collected from a finite resource in order of desirability until the yield of a patch is poor enough to warrant the time investment to travel to another.  The analogy is completed when we consider that it takes considerable time to design and construct missions.  Though different missions are always being run in parallel, they consume a fixed portion of the budget which could otherwise have gone to some other mission.
	
	We will consider a mission at fixed cost that takes a development period $\tdev$ before it begins collecting data.  Then the yield as a function of time is $N(t)=N_0(t-\tdev)^{q_t}$, where it should be understood that when this quantity is not positive, it is 0.  It is convenient to nondimensionalize the variables in this expression by measuring time in years, so that $N_0$ represents the number of planets returned in the first year of operation.  To determine the optimal stopping time we use the marginal value theorem, which maximises the average information gain per time, this sets $\dot S(t_0) = S(t_0)/t_0$.  This can then be solved for $t_0$, yielding 
	\beq
	t_{operation} = \frac{1}{W_0\left(N_0^{1/q_t}\tdev/e\right)}\tdev
	\eeq
	Here, $W_0(x)$ is the Lambert W function, which is the solution to the equation $W_0e^{W_0}=x$, and asymptotes to a logarithm for large arguments.  So, it can be seen that the operation time scales not quite linearly with development time, and decreases with increasing first year yield.  The total science return over the mission lifetime is given by
	\beq
	S(t_0) = q_t\left[W_0\left(N_0^{1/q_t}\tdev/e\right)+1\right]=q_t\left[\frac{t_{dev}}{t_{op}}+1\right]
	\eeq

\subsection{When is it best to follow a prediction, and when is it best to challenge it?}\label{challenge}
	
	We now turn our attention to the scenario where, according to some leading model of habitability, a particular locale should be devoid of life.  According to this model, any attempt at measuring a biosignature around such a location would be fruitless, and would have been better spent collecting data from a safer bet.  However, we are unlikely to ever be $100\%$ confident in any of our habitability models, and so it is possible that testing the model will result in it being overturned.  Here, we show that our framework allows us to determine a criterion for when this habitability model should be obeyed, and when it should be challenged, based off of the confidence one assigns to its truthfulness.  A concrete example would be whether to check hot Jupiters for biosignatures, as by most reasonable accounts they do not fulfil the requirements for life \citep{fortney2021hot}.
	
	As a toy example, let us consider the case where there are two possibilities: the leading theory, which dictates that life is impossible on some type of system, and a second theory, which assigns a nonzero chance that life may exist.  The confidence we assign to the standard theory is denoted $p$.  The prior distribution for the fraction of systems harbouring a signal is then
	\beq
	p_b(f)=p\,\delta(f)+(1-p)\,\delta(f-f_2)
	\eeq
	Here $\delta$ is the delta function, which equals 0 when its argument is nonzero and integrates to 1.  More thoroughly, we could consider a range of values for $f_2$, but this significantly complicates the analysis, and leads to little new insight.
	
	Now we determine the expected amount of information gained from a measurement of a system.  According to the above prior, the probability of measuring no signal is $P_0=p+(1-p)(1-f_2)$, and the probability of measuring a signal is $P_1=(1-p)f_2$.  Then the expected information gain $\langle\Delta S\rangle = P_0 S(p(f|\text{no detection}))+P_1S(p(f|\text{detection}))-S(p_b(f))$ is:
	\bea
	\langle\Delta S\rangle=(1-p)(1-f_2)\log\left(\frac{(1-p)(1-f_2)}{p+(1-p)(1-f_2)}\right)\nonumber\\
 -p\log(p+(1-p)(1-f_2))-(1-p)\log(1-p)\label{f10}
	\eea
	If we further take the limits $p\rightarrow1$, $f_2\rightarrow0$, this expression simplifies somewhat:
	\beq
	\langle\Delta S\rangle\rightarrow(1-p)\Big[f_2(1-\log(1-p))+(1-f_2)\log(1-f_2)\Big]
	\eeq
	From this, it can be seen that the expected amount of information gain is proportional to $(1-p)$, the probability that the theory's recommendation is incorrect.
	
	\begin{centering}
		\begin{figure}
			\centering
			\includegraphics[width=.5\textwidth]{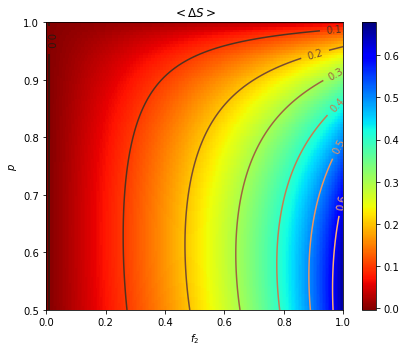}
			\caption{Information gain for measurement of an unlikely signal as a function of possible signal occurrence rate $f_2$ and confidence in the theory proclaiming its impossibility $p$.}
			\label{pf2}
		\end{figure}
	\end{centering}
	
	Eqn. \ref{f10} is displayed in Fig. \ref{pf2}, from where it can be seen that expected information is maximised when $f_2$ is large and $p$ is small.  As example values, take $\langle\Delta S(p=.9,f_2=.1)\rangle=.02$, $\langle\Delta S(p=.99,f_2=.1)\rangle=.005$.
	
	To determine whether this measurement should be made, we need to compare it to an alternative measurement yielding a different amount of expected information gain, and see which is larger.  In the standard case of measuring a biosignature of frequency $f$ well localised from 0, the information is $S=1/2\log(f(1-f)/N)$ after $N$ measurements, and so the information gain is $\Delta S = 1/(2N)$.  So, for the case $p=.99$, $f_2=.1$, the expected value for the risky measurement exceeds the standard after the standard measurement has been made 107 times.  This suggests a strategy of collecting measurements around likely places first, until the gains become small enough that measuring the unlikely signal becomes more interesting, on the off chance that it could invalidate the current theory of habitability.
	
	The above assumed that no measurement of the unlikely signal had been attempted previously.  We may wish to determine how many times to challenge a theory before accepting it.  For this scenario, if $N$ previous measurements have been made, the prior will be
	\beq
	p_b(f)\propto p\,\delta(f)+(1-p)(1-f_2)^N\delta(f-f_2)
	\eeq
	With constant of proportionality enforcing that this is normalised to 1.  The expected information gain is, in the limit of  $N\rightarrow\infty$, 
	\beq
	\langle\Delta S\rangle\rightarrow \frac{1-p}{p}\,N\,f_2\,(1-f_2)^N\,\log\left(\frac{1}{1-f_2}\right)
	\eeq
	
	\begin{centering}
	\begin{figure}
		\centering
		\includegraphics[width=.45\textwidth]{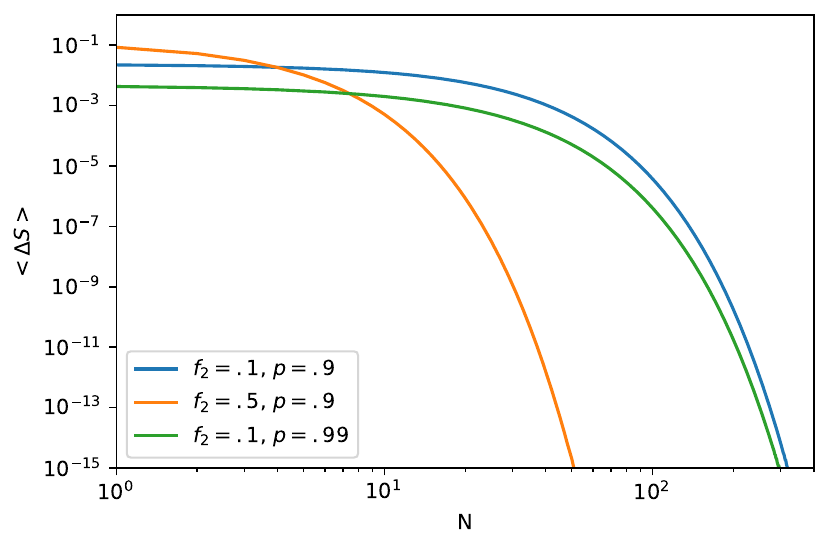}
		\caption{Information gain of an unlikely signal after a series of $N$ unsuccessful measurements.  Above $N\gtrsim1/f_2$ the likelihood of measuring a signal becomes exponentially suppressed.}
		\label{pf2N}
	\end{figure}
	\end{centering}

	The full expression is plotted in Fig. \ref{pf2N}.  From here, we can clearly see a steep dropoff at $N>1/f_2$, beyond which empirical evidence strongly indicates the absence of a signal.

\section{Discussion}

In this work, we have utilised the concept of \emph{information gain} as a method to assess the value of biosignature missions, and demonstrated that this is capable of  generating nontrivial recommendations for mission design and target prioritisation, across various domains.  Throughout, this has resembled less of a first-principles approach, whereby any mission setup can automatically be loaded into a master equation to straightforwardly generate recommendations, but rather a style of thinking about missions in terms of the quality of the knowledge that they generate as a function of mission parameters.  The concept of information explicitly depends on the quantities that are abstracted out of the raw data any mission would return, and so at the outset any analysis of this sort relies on value judgements on the part of the practitioner.  We have illustrated that this process does not always yield a unique encapsulation of the data, as for instance when we chose to define how we detect population trends through the distribution of the occurrence rate ratio, rather than some other measure.  While there is no guarantee that the recommendations will be independent of the way we set up the problem, the advantage this framework has is in making the mission goals and assumptions explicit when deriving recommendations.  

At this point we'd like to reiterate that we are ardently against using information gain as a quantity to maximise for designing missions, since attempts to substitute quantitative proxies for vague overarching goals often results in pathological maximisation procedures, often to the detriment of the original intent \citep{chrystal2003goodhart}.  Rather, we advocate the use of this framework for generating rules of thumb and an intuitive understanding for when certain mission aspects ought to be improved upon.  Additionally, we note that this framework explicitly depends upon an honest and accurate assessment of mission capabilities.  Even with rigorous design practices, often there ends up being uncertainties in instrumental sensitivities, background noise, and plausibility of modelling scenarios that prevent perfect assessment of the relevant parameters.  On top of this, human biases preclude even perfect experts from being able to make impartial assessments for any new venture \citep{pohorille2020evaluating}.  Nevertheless, our formalism is useful for determining the sensitivity of our recommendations to these unknowns.

We hope that by demonstrating the practicality of this framework to a variety of cases, the reader will find it productive to apply to other, new problems.  To this end, our approach has often been the following: first, we seek out a continuous quantity that can be measured with a particular mission setup.  Next, we compute the distribution of values compatible with observations, given the model parameters and any other additional assumptions that must be used.  Then, lastly, we vary these mission parameters, subject to any relevant mission constraints such as finite budget/lifetime/sample size, and determine the combination of mission parameters that equivalently maximizes information, minimizes uncertainty or maximizes probability of the desired outcome.  Our hope is that this may be useful as a tool to aid in determining the design parameters of future missions, providing transparency and impartiality in selecting among upcoming missions, and ultimately maximizing our chances of successfully detecting biosignatures.\\

\noindent{\bf Data Availability.} Code and data used in this paper are located here: \url{github.com/SennDegrare/Bio-signatures}.\\

\noindent{\bf Declaration of Interests.} The authors report no conflict of interest.
\ack[Acknowledgements]{We would like to thank Jacob Haqq Misra, Joe Lazio and Joe Silk for fruitful discussions.}

\bibliographystyle{apalike}
\bibliography{bibliography} 




\end{document}